\documentclass[prl,twocolumn,superscriptaddress,floatfix,showpacs,aps,final]{revtex4}
\usepackage{graphicx} 
\usepackage{color} 
\usepackage{amsmath} 
\usepackage[urlcolor=blue, hyperindex, colorlinks, bookmarks=true,linkcolor=black,citecolor=black]{hyperref} 
\usepackage{amssymb}
\usepackage{psfrag}

\newcommand{\dg}{^\dagger}

\newcommand{\bra}[1]{\langle{#1}|}
\newcommand{\ket}[1]{|{#1}\rangle}

\DeclareMathOperator{\rre}{Re}
\DeclareMathOperator{\iim}{Im}

\begin{document}

\title{Randomized benchmarking and process tomography for gate errors in a solid-state qubit}
\date{November 26, 2008}
\author{J. M. Chow}
\affiliation{Departments of Physics and Applied Physics, Yale University, New Haven, Connecticut 06520, USA}
\author{J. M. Gambetta}
\affiliation{Institute for Quantum Computing and Department of Physics and Astronomy, University of Waterloo, Waterloo, Ontario, Canada N2L 3G1}
\author{L. Tornberg}
\affiliation{Chalmers University of Technology, SE-41296 Gothenburg, Sweden}
\author{Jens Koch}
\affiliation{Departments of Physics and Applied Physics, Yale University, New Haven, Connecticut 06520, USA}
\author{Lev S. Bishop}
\affiliation{Departments of Physics and Applied Physics, Yale University, New Haven, Connecticut 06520, USA}
\author{A. A. Houck}
\affiliation{Departments of Physics and Applied Physics, Yale University, New Haven, Connecticut 06520, USA}
\author{B. R. Johnson}
\affiliation{Departments of Physics and Applied Physics, Yale University, New Haven, Connecticut 06520, USA}
\author{L. Frunzio}
\affiliation{Departments of Physics and Applied Physics, Yale University, New Haven, Connecticut 06520, USA}
\author{S. M. Girvin}
\affiliation{Departments of Physics and Applied Physics, Yale University, New Haven, Connecticut 06520, USA}
\author{R. J. Schoelkopf}
\affiliation{Departments of Physics and Applied Physics, Yale University, New Haven, Connecticut 06520, USA}

\begin{abstract}
We present measurements of single-qubit gate errors for a superconducting qubit. Results from  quantum process tomography and randomized benchmarking are compared with gate errors obtained from a double $\pi$ pulse experiment.  Randomized benchmarking reveals a minimum average gate error of $1.1\pm0.3\%$ and a simple exponential dependence of fidelity on the number of gates. It shows that the limits on gate fidelity are primarily imposed by qubit decoherence, in agreement with theory.  
\end{abstract}
\maketitle

The success of any computational architecture depends on the ability to perform a large number of gates, and gate errors meeting a fault-tolerant threshold. While classical computers today perform many operations without the need for error correction, gate error thresholds for quantum error correction are still very stringent, with conservative estimates on the order of $10^{-4}$  \cite{gottesman, knill_nature_2005}.

Gate fidelity is the standard measure of agreement between an ideal operation and its experimental realization. Beyond the gate fidelity, identifying the nature of the dominant errors in a specific architecture is particularly important for improving performance. While NMR, linear optics, and trapped ion systems are primarily limited by systematic errors such as spatial inhomogeneities and imperfect calibration \cite{childs_tomo_2001, obrien_QPT_2004, riebe_tomo_2006}, for solid-state systems decoherence is the limiting factor. The question of how to measure gate errors and distinguish between various error mechanisms has produced different experimental metrics for gate fidelity, such as the double $\pi$ metric employed in superconducting qubits \cite{lucero_gates_2008}, process tomography as demonstrated in trapped ions, NMR, and superconducting systems \cite{obrien_QPT_2004, riebe_tomo_2006, childs_tomo_2001, neeley}, and randomized benchmarking, as performed in trapped ions and NMR \cite{knill_randomized_2008, ryan_randomized_2008}.

Here we present measurements of single-qubit gate fidelities where the three metrics mentioned above are implemented in a circuit QED system \cite{wallra_strong_2004, blais_cavity_2004} with a transmon qubit \cite{koch_charge-insensitive_2007}. We compare the results for the different metrics and discuss their respective advantages and disadvantages. We find single-qubit gate errors at the $1$$\sim$$2\%$ level consistently among all metrics. These low gate errors reflect recent improvements in coherence times \cite{schreier_suppressing_2008, houck_purcell_2008}, systematic microwave pulse calibration, and accurate determination of gate errors despite limited measurement fidelity. In circuit QED, measurement fidelity can be as high as 70\%, though in this experiment it is $\sim$$5\%$, as readout is not optimized. The magnitude of errors and their dependence on pulse length are consistent with the theoretical limits imposed by qubit relaxation and the presence of higher qubit energy levels, with only small contributions from calibration errors. 
 
We first discuss the double $\pi$ metric ($\pi$-$\pi$). Here, two $\pi$ pulses are applied in succession, which ideally should correspond to the identity operation $\openone$. The aim of $\pi$-$\pi$ is to determine the deviations from $\openone$ by measuring the residual population of the excited state following the pulses. Despite its simplicity, this metric captures the effects of qubit relaxation and the existence of levels beyond a two-level Hilbert space. However, in general, it is merely a rough estimate of the actual gate fidelity as it does not contain information about all possible errors. In particular, errors that affect only eigenstates of $\sigma_x$ or $\sigma_y$ and deviations of the rotation angle from $\pi$ are not well captured by this measure. 

A second metric that, in principle, completely reveals the nature of all deviations from the ideal gate operation is quantum process tomography (QPT) \cite{chuang_blackbox_1997}. Ideally, QPT makes it possible to associate deviations with specific error sources, such as decoherence effects or non-ideal gate pulse calibration. However, in systems with imperfect measurement, it is difficult to assign the results from QPT to a single gate error.  Moreover, the number of measurements that are necessary for QPT scales exponentially with the number of qubits.

	While QPT provides information about a single gate, randomized benchmarking (RB) \cite{Emerson:2007a, knill_randomized_2008} gives a measure of the accumulated error over a long sequence of gates. This metric hypothesizes that with a sequence of randomly chosen Clifford group generators ($R_{u}=e^{\pm i\sigma_u \pi/4}$, $u=x,y$) the noise can behave as a depolarizing channel, such that an average gate fidelity can be obtained. In contrast to both $\pi$-$\pi$ and QPT, RB is approximately independent of errors in the state preparation and measurement. Also, while the other metrics measure a single operation and extrapolate the performance of a real quantum computation, RB tests the concatenation of many operations (here up to $\sim200$), just as would be required in a real quantum algorithm.

\begin{figure}[t]
\centering
\includegraphics[scale=1]{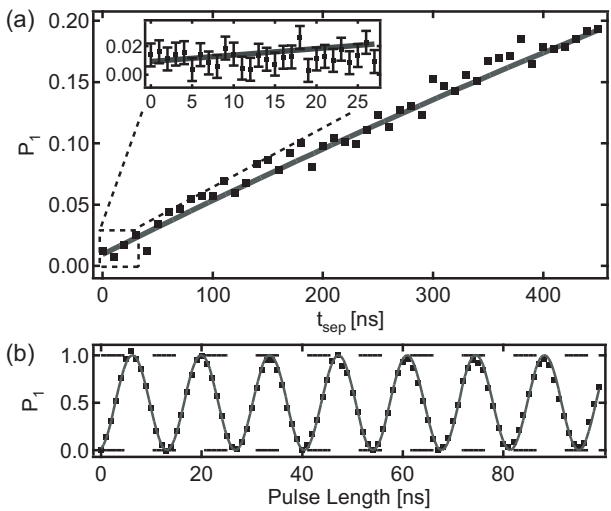}
\caption{(a)~Excited state qubit population $P_{1}$ vs. separation time $t_{\text{sep}}$ between two successive $\pi$-pulses ($\sigma = 2$\,ns). The data agree well with the simulation (solid line) involving relaxation and decoherence. The inset shows additional data taken for $0 \leq t_{\text{sep}} \leq 30\,\text{ns}$. The residual population corresponding to the minimal separation is found to be $0.014 \pm 0.008$ giving a single qubit gate error of $0.7\pm 0.4 \%$. (b)~Rabi oscillations show a  visibility of $100.4 \pm 1.0 \%$.\label{figure1}} 
\end{figure}

The gate error metrics are performed in a circuit QED sample consisting of a transmon qubit coupled to a coplanar waveguide resonator \cite{wallra_strong_2004,blais_cavity_2004, koch_charge-insensitive_2007}. The theory and discussion, however, extend generally to all qubit systems including ions and spins.  The sample fabrication and measurement techniques are similar to those in Refs.\ \cite{majer_coupling_2007,schreier_suppressing_2008,houck_purcell_2008}. Experimentally measured parameters include the qubit-cavity coupling strength given by $g_0/\pi = 94.4$ MHz, the resonator frequency $\omega_r/2\pi= 6.92$ GHz, photon decay rate of $\kappa/2\pi = 300$ kHz, and qubit charging energy $E_C/2\pi=340$\,MHz. The qubit is detuned from its flux sweet spot by $\sim1.5$\,GHz with a resonant frequency of $\omega_{01}/2\pi = 5.96$ GHz, and coherence times of $T_1 = 2.2$ $\mu$s and $T_2^* = 1.3$ $\mu$s.

	In analogy to the NMR language, our single-qubit operations are rotations about the $x-$, $y-$, and $z-$axes of the Bloch sphere \cite{slichter}. Rotations about any axis in the $x$-$y$ plane are performed using microwave pulses. The carrier frequency is resonant with the qubit transition frequency and the pulse amplitudes and phases define the rotation angle and axis orientation, respectively. In all experiments, the pulse-shape is Gaussian with standard deviation $\sigma$ between 1 and 12\,ns. The pulses are truncated at $2\sigma$ on each side and a constant buffer time of 8\,ns is inserted after each pulse to ensure complete separation of the pulses. Using tune-up sequences similar to those used in NMR \cite{calibration_NMR}, each pulse amplitude is calibrated by repeated application of the pulse and matching the measurement outcome to theory. (See supplementary material for details.)

\emph{Double $\pi$}.---After calibration, we perform the $\pi$-$\pi$ experiments with $\sigma$ = 2\,ns and varying separation time $t_{\text{sep}}$ between the two $\pi$ gates. Subsequently, the excited state probability $P_{1}$ is measured, as shown in Fig.\ 1(a). Due to the decay of the excited state following the first $\pi$ pulse, $P_{1}$ increases as a function of $t_{\text{sep}}$. This can be accurately captured in simulations with a simple theoretical model consisting of the dynamics from a master equation for a driven three-level atom subject to relaxation and dephasing, with corresponding time-scales $T_1$ and $T_{\phi}$. The coherent evolution is governed by the Hamiltonian 
	\begin{equation}
		H=\hbar \sum_{j=1,2} \left[\omega_{0j}\sigma_j^{\dag}\sigma_j +\varepsilon_j(t)(\sigma_j^{\dag}+\sigma_j)\right],
	\end{equation}
where $\sigma_j = \ket{j-1}\bra{j}$ is the lowering operator for the multi-level atom with eigenenergies $\hbar\omega_j$. The corresponding transition energies are denoted $\hbar\omega_{ij} =\hbar(\omega_j - \omega_i)$. Drive strength and pulse-shapes are determined by 
	\begin{equation}
		\varepsilon_j(t) = \frac{ g_j^2}{\omega_r-\omega_{j-1,j}}\left[X(t)\cos(\omega_d t) + Y(t) \sin (\omega_d t) \right].
	\end{equation} 
Here, $g_j\sim\sqrt{j}g_0$ is the transmon coupling strength \cite{koch_charge-insensitive_2007}, $\omega_d/2\pi$ is the frequency of the drive, and $X(t)$ and $Y(t)$ are the pulse envelopes in the two quadratures. 
	
	The inset of Fig.\,1(a) shows the experiment with $t_{\text{sep}}$ varying between 0\,ns and 30\,ns repeated $2.5 \times 10^6$ times. We measure $P_1=0.014\pm0.008$ at $t_{\text{sep}} = 0$\,ns. Dividing this probability by two as in Ref.\,\cite{lucero_gates_2008} gives a single gate error of $0.7\pm0.4\%$. 

	Conceptually, the $\pi$-$\pi$ measure is similar to the visibility measure used by Wallraff \emph{et al.\,}in Ref.\,\cite{wallra_approaching_2005}, corresponding to $(1-\langle\sigma_z\rangle)/2$ after a single $\pi$ pulse.  Fig.\,1(b) shows Rabi oscillations made by increasing the length of a pulse resonant with the qubit transition frequency. The visibility is found to be $100.4\pm1.0\%$. This also agrees with our simple theoretical model taking into account the $T_1$, $T_2$, and third-level at our specific operating point. 
	
\begin{figure}[htbp]
\centering
		    \psfrag{F}[c][][1.0]{$\rre \chi$}
		    \psfrag{G}[c][][1.0]{$\iim \chi$}
		    \psfrag{R}[l][][1.0]{$\openone$}
		    \psfrag{S}[l][][1.0]{$R_x(\pi/2)$}
		    \psfrag{T}[l][][1.0]{$R_y(\pi/2)$}
		    \psfrag{I}[l][][0.8]{$\openone$}
		    \psfrag{X}[l][][0.8]{$\sigma_x$}
		    \psfrag{Y}[l][][0.8]{$\sigma_y$}
		    \psfrag{Z}[l][][0.8]{$\sigma_z$}
\includegraphics[scale=0.97]{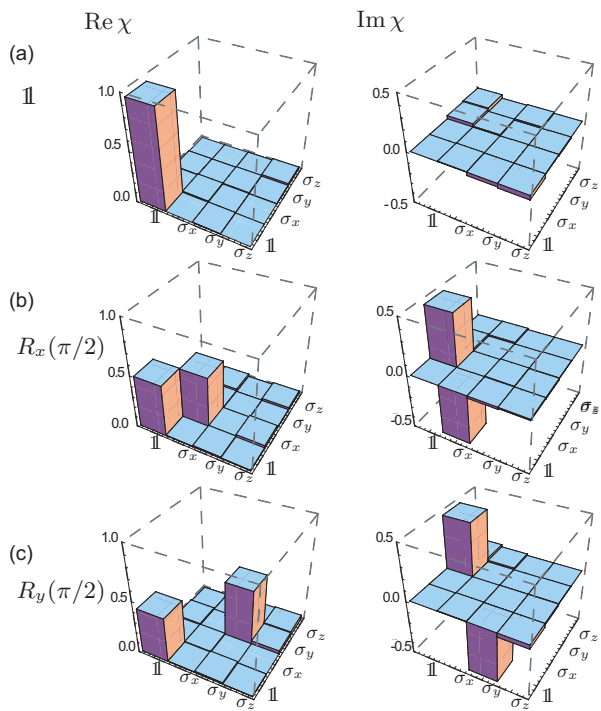}
\caption{(color online) Real and imaginary parts of the experimentally obtained process matrix $\chi$ for the three processes (a) $\openone$, (b) $R_x(\pi/2)$, and (c) $R_y(\pi/2)$ for $\sigma = 2$\,ns. 
\label{figure2}}
\end{figure}	
		
\emph{Quantum Process Tomography}.---The idea behind QPT is to determine the completely positive map $\mathcal{E}$, which represents the process acting on an arbitrary input state $\rho$. The theory is detailed in Refs.\,\cite{poyatos_process_1997,chuang_blackbox_1997} and can be summarized as follows. Any process for a $d$ dimensional system (for 1 qubit $d =2$) can be written as 
\begin{equation}\label{eq:map}
  \mathcal{E}(\rho)=\sum_{m,n=0}^{d^2-1} \chi_{mn} B_m \rho B_n\dg
\end{equation} where $\{B_n\}$ are operators which form a basis in the space of $d\times d$ matrices, and $\chi$ is the process matrix. To determine $\chi$, we prepare $d^2$ linearly independent input states  $\{\rho^\mathrm{in}_n\}$.  For every input state, the output state  $\rho^\mathrm{out}_n=\mathcal{E}(\rho^\mathrm{in}_n)$  is determined by state tomography. The process matrix is then obtained by inverting Eq.\,\eqref{eq:map}. However, in general this last step does not guarantee a completely positive map. To remedy this, we use a maximum likelihood estimation based on Ref.\,\cite{obrien_QPT_2004}, which is detailed in the supplementary material.

We perform QPT on the three processes $\openone$, $R_x(\pi/2)$ and $R_y(\pi/2)$ using the four linearly independent input states $\ket{0}, \ket{1}, (\ket{0}+i\ket{1})/\sqrt{2}$, and $(\ket{0}-\ket{1})/\sqrt{2}$. The results of this procedure are shown in Fig.\,2. Here, bar plots of the real and imaginary parts of $\chi$ are shown for a pulse with $\sigma=2$\,ns in the Pauli basis  $\{B_n\}= \{\openone, \sigma_x, \sigma_y,\sigma_z\}$. We can compare our data to the ideal process matrices $\chi_{\text{ideal}}$. For instance, for the $\openone$ process, we expect $\chi_{\openone\openone} =1$ and $\chi_{uu'}=0$ otherwise, which is in good agreement with the measured results. Small deviations from $\chi_\text{ideal}$ arise from preparation and measurement errors, gate over-rotations, decoherence processes, qubit anharmonicity, etc. Calibration errors of the pulses in the $x$ axis are seen as a non-zero $\iim\{\chi_{\openone \sigma_x}\}$ and a drive detuning error is exhibited in $\iim\{\chi_{\openone \sigma_z}\}$.

From the experimentally obtained process matrix $\chi$ and its ideal counterpart $\chi_{\mathrm{ideal}}$ we can directly calculate the process fidelity, $F_p=\mathrm{Tr}[\chi_{\mathrm{ideal}}\chi]$, and the gate fidelity $F_g = \int d\psi \bra{\psi}U^{\dag}\mathcal{E}(\psi)U\ket{\psi}$. Here the integral uses the uniform measure $d\psi$ on the state space, normalized such that $\int d\psi =1$.  $F_g$ can be understood as how close $\mathcal{E}$ comes to the implementation of the unitary $U$ when averaged over all possible input states $\ket{\psi}$. From Ref.\,\cite{horodecki, nielsen_gatefid_2002}, there is a simple relationship between the $F_p$ and $F_g$, namely $F_g = (d F_p +1)/(1+d)$. For the three processes displayed in Fig.\,2, $F_p$ is 0.96, 0.95, and 0.95 $\pm0.01$.

\begin{figure}[t]
\centering
\psfrag{d}[l][][1.0]{$(1-F_g)$}
\psfrag{I}[l][][1.0]{$\openone$}
\psfrag{R}[l][][1.0]{$R_x(\pi/2)$}
\psfrag{Q}[l][][1.0]{$R_y(\pi/2)$}
\includegraphics[scale=0.97]{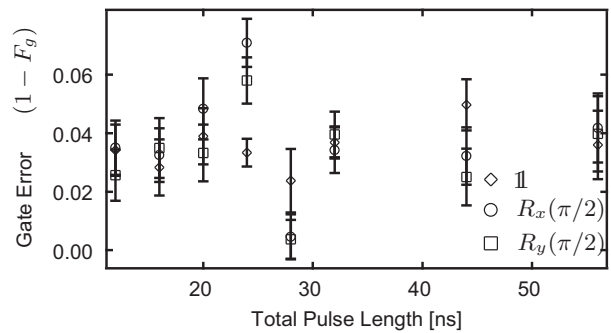}
\caption{Gate error vs. total pulse length obtained from quantum process tomography plotted for the three processes $\openone, R_x(\pi/2), R_y(\pi/2)$.\label{figure3}}
\end{figure}

Figure 3 shows $F_g$ as a function of pulse length. The error bars are standard deviations obtained by repeating the maximum-likelihood estimation for input values chosen from a distribution with mean and variance given by measurement. The missing increase of gate error with pulse length is currently not well understood and may be partly due to errors in the measurement and preparation.

\emph{Randomized Benchmarking}.---The RB protocol, described in Knill \emph{et al.\,}\cite{knill_randomized_2008}, consists of the following: (1) initialize the system in the ground state, (2) apply a sequence of randomly chosen pulses in the pattern $\prod_i C_i P_i$ where $C_i$ are Clifford group generators $e^{\pm i \sigma_u \pi/4}$, with $u = x,y$, and $P_i$ are Pauli rotations, i.e. $\openone, \sigma_x, \sigma_y, \sigma_z$, (3) apply a final Clifford or Pauli pulse to return to one of the eigenstates of $\sigma_z$, (4) perform repeated measurements of $\sigma_z$, and compare to theory to obtain the sequence fidelity.

\begin{figure}[t]
\centering
\includegraphics[scale=0.97]{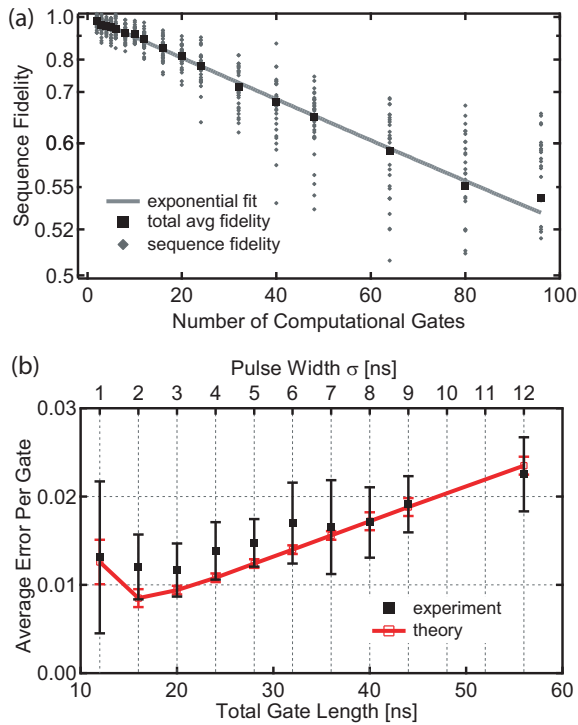}
\caption{(color online) (a)~Average fidelity vs.\,number of applied computational gates. Computational gates consist of a randomized Pauli with a randomized Clifford generator. For $\sigma$ of 3\,ns we obtain an average gate error of 1.1$\%$. (b)~Average error per gate (experimental and theoretical) at different pulse widths. The rise for $\sigma < 2$\,ns corresponds to the onset of limitation by the third level of the transmon. The increase in error per gate for $\sigma > 2$\,ns is due to the limitation by relaxation. \label{figure4}}
\end{figure}
We choose the number of randomizations, sequences, and sequence lengths exactly as in Ref.\,\cite{knill_randomized_2008} with the longest sequences consisting of 196 pulses. All 544 final pulse sequences are applied for 250,000 measurements each, taking a total time of about an hour.

	The average fidelity is an exponentially decaying function with respect to the number of gates. Fig.\,4(a) plots the fidelity as a function of the number of computational gates for all randomized sequences with $\sigma$\,=\,3\,ns. An average error per gate of $0.011\pm 0.003$ is obtained by averaging over all the randomizations and fitting to the exponential decay. The excellent fit to a single exponential indicates a constant error per gate, consistent with uncorrelated random gate errors due to $T_1$, $T_\phi$, and no other mechanisms significantly affecting repeated application of single-qubit gates. The reduction of the error by a factor of $\sim$$1/3$ from QPT is likely due to the over-estimation of errors in QPT where gate errors cannot be isolated from measurement and preparation errors.

	The benchmarking protocol is repeated for different pulse widths $\sigma$, and the average error per gate is extracted, plotted versus total gate length, and compared to theory in Fig.\,4(b). At large gate lengths, experimental results agree well with theory. In this regime, errors are dominated by relaxation and dephasing. At small gate lengths, the gate fidelity is limited by the finite anharmonicity and the resulting occupation of the third level. We obtain error bars from standard deviations in error per gate having generated fidelity values from distributions with means and variance obtained from the experiment and theory. The optimal gate length is found to be 20\,ns, as shown in Fig.\,4(b). 

		\begin{table}[b!]
	\centering
		\begin{tabular}{lc}\hline\hline
		Metric & Measured error in $\%$ \\\hline
		$\pi$--$\pi$ &		$0.7\pm0.4$		 \\
		Process tomography: $\openone$&  $2.4\pm1.1$     \\
		Process tomography: $R_x(\pi/2)$& $2.6\pm0.8$    \\
		Process tomography: $R_y(\pi/2)$& $2.2\pm0.7$    \\
		Randomized benchmarking & $1.1\pm0.3$\\\hline\hline
		\end{tabular}
	\caption{Gate errors for the three metrics used in this work. The measurements show consistently low gate errors of the order of $1\sim2\%$.\label{tab:summary}}
\end{table}

	\emph{Conclusions}.---We have systematically investigated gate errors in a circuit QED system by measuring gate fidelity using the $\pi$-$\pi$ metric, quantum process tomography, and randomized benchmarking. Table I summarizes our results and displays consistently low gate errors across all metrics. From comparison with theory, we conclude that the observed magnitude of errors fully agrees with the limitations imposed by qubit decoherence and finite anharmonicity. Specifically, in the $T_1$ limited case and for moderate gate lengths $t_g$, we find that the gate error scales as $\sim t_g /T_1$. Once coherence times of superconducting qubits and pulse-shaping are improved, the aforementioned metrics will be useful tools for characterizing gate fidelities as they approach the fault-tolerant threshold. Randomized benchmarking will be a particularly attractive option for multi-qubit systems due to its favorable scaling properties as compared to QPT.
	
\begin{acknowledgements}
We acknowledge E. Knill, R. Laflamme, K. Resch, and C. Ryan for valuable discussions. This work was supported by NSA under ARO contract W911NF-05-1-0365 and by the NSF under grants DMR-0653377 and DMR-0603369. JMG was supported by CIFAR, MITACS, and ORDCF.  LT was supported by the EU through IST-015708 EuroSQIP and by the SRC.  
\end{acknowledgements}

\end{document}


\title{Supplementary material for ``Randomized benchmarking and process tomography for gate errors in a solid-state qubit"}

\date{November 26, 2008}
\author{J. M. Chow}
\affiliation{Departments of Physics and Applied Physics, Yale University, New Haven, Connecticut 06520, USA}
\author{J. M. Gambetta}
\affiliation{Institute for Quantum Computing and Department of Physics and Astronomy, University of Waterloo, Waterloo, Ontario, Canada N2L 3G1}
\author{L. Tornberg}
\affiliation{Chalmers University of Technology, SE-41296 Gothenburg, Sweden}
\author{Jens Koch}
\affiliation{Departments of Physics and Applied Physics, Yale University, New Haven, Connecticut 06520, USA}
\author{Lev S. Bishop}
\affiliation{Departments of Physics and Applied Physics, Yale University, New Haven, Connecticut 06520, USA}
\author{A. A. Houck}
\affiliation{Departments of Physics and Applied Physics, Yale University, New Haven, Connecticut 06520, USA}
\author{B. R. Johnson}
\affiliation{Departments of Physics and Applied Physics, Yale University, New Haven, Connecticut 06520, USA}
\author{L. Frunzio}
\affiliation{Departments of Physics and Applied Physics, Yale University, New Haven, Connecticut 06520, USA}
\author{S. M. Girvin}
\affiliation{Departments of Physics and Applied Physics, Yale University, New Haven, Connecticut 06520, USA}
\author{R. J. Schoelkopf}
\affiliation{Departments of Physics and Applied Physics, Yale University, New Haven, Connecticut 06520, USA}

\maketitle

\section{Microwave Pulse-Shaping and Concatenation}

 Fig.\,S1(a) shows a measured single microwave qubit control pulse as used in our experiments. The Gaussian pulse envelopes are generated with a 10-bit Tektronix AWG520 arbitrary waveform generator with 1\,ns resolution. The pulse shapes are mixed with sine and cosine waves at the qubit transition frequency of $5.95$\,GHz using an Agilent E8267 Vector Signal generator. It is thus possible to produce pulses phase-shifted by $90^{\circ}$ for qubit rotations around $x$ and $y$. Rotations about the $z$-axis were performed with a rotation of the reference frame with an accompanying delay equivalent to the $x$ and $y$ pulses, see Ref.\,\cite{knill_randomized_2008}. For example, the sequence $R_x(\pi) R_z(\pi) R_y(\pi)$ becomes $R_x(\pi) \openone R_y(-\pi)$.
 
  The inclusion of the relatively long buffer time of 8\,ns at the end of each pulse is to ensure that a single microwave pulse is completely turned off before the next one is applied. With sequences of up to $\sim$$200$ concatenated pulses in the randomized benchmarking experiments, avoiding residual pulse overlap becomes very important. Fig.\,S1(b) is a sample sequence for randomized benchmarking measured on a fast-sampling scope, and corresponds to applying $R_y(\pi)R_y(-\pi/2)R_z(-\pi)R_x(\pi/2)...$ $R_y(-\pi/2)R_x(\pi)R_z(\pi)R_y(-\pi)$.
 
 \begin{figure}[t]
\centering
\includegraphics[scale=1]{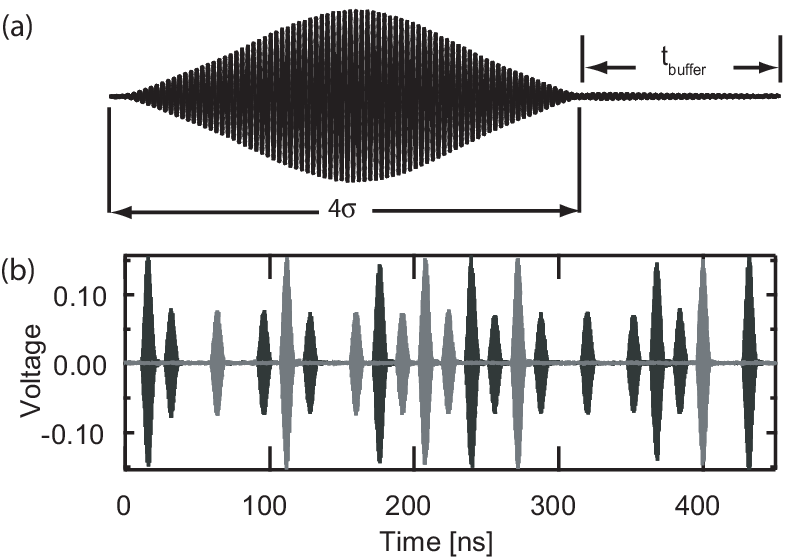}
\caption{(a)~Measured single microwave qubit control pulse. (b)~Sample randomized benchmarking sequence. Pulses in black (grey) correspond to rotations in the $y(x)$-axis\label{figureS1}} 
\end{figure}

\section{Qubit Pulse Calibration} 
  In order to obtain the best gate fidelity, accurate pulse amplitude calibration is necessary. We perform some tune-up schemes similar to those used in NMR experiments \cite{calibration_NMR}, which involve applying repeated pulses such that small errors can be built up. The rotations $R_x(\pm\pi/2)$, $R_y(\pm\pi/2)$, $R_x(\pm\pi)$, $R_y(\pm\pi)$ are all calibrated independently. 
 
 For the $\pi/2$ rotation around the $x$-axis, we apply sequences consisting of an odd number $n$ of the pulses, $[R_x(\pi/2)]^n$, and subsequently measure $\langle \sigma_z\rangle$. Performing this for several $n$, we calibrate the amplitude by comparing the measured values with the ideal outcomes obtained from a simple two-level simulation including decoherence. The other $\pi/2$ pulses are then calibrated similarly. 
 
 Once all $\pi/2$ pulses are calibrated, the $\pi$ pulses are calibrated with a slightly different scheme. To calibrate a $\pi$ pulse around the $x$-axis, we first apply a single calibrated $R_x(\pi/2)$ pulse followed by an integer number $m$ of the $\pi$ pulses, $R_x(\pi/2)[R_x(\pi)]^m$. Again, the measured values of $\langle \sigma_z\rangle$ for different $m$ are compared with simulation to calibrate the amplitude of the $R_x(\pi)$ pulse. Note that the $\pi$ pulse calibration contains a prepended $\pi/2$ pulse in order to render the scheme first-order sensitive to deviations in the amplitude. The scheme is then repeated for the other $\pi$ pulses as well. 
 
\section{QPT pulse sequences and maximum likelihood estimation}
 In all the QPT experiments, the measurements are performed after sequences of three concatenated pulses are applied to the qubit. The first pulse, chosen from $\{\openone, R_{x}(\pi), R_x(\pi/2), R_y(\pi/2)\}$, prepares the four linearly independent input states $\ket{0}, \ket{1}, (\ket{0}+i\ket{1})/\sqrt{2}$, and $(\ket{0}-\ket{1})/\sqrt{2}$, whose projectors span the space of  $2\times2$ density matrices $\rho$. The second pulse corresponds to the investigated process chosen from, $\{\openone, R_x(\pi/2), R_y(\pi/2)\}$. A final pulse ($\{\openone, R_{x}(\pi), R_x(\pi/2), R_y(\pi/2)\}$) rotates the measurement axis to perform state tomography on the state resulting from the first two pulses. 
 
The state tomography data allows us to construct the process $\mathcal{E}(\rho)$, which can be written as \cite{chuang_blackbox_1997}
\begin{equation}
\mathcal{E}(\rho) = \sum_{mn} \chi_{mn}B_m\rho B_n\dg
\end{equation} where $\chi_{mn}$ is known as the process matrix and is defined with respect to an operator basis, which we take as the Pauli basis $\{B_n\}=\{\openone, \sigma_x,\sigma_y,\sigma_z\}$. By definition the $\chi$ matrix must be Hermitian and due to the completeness constraint must satisfy \cite{chuang_blackbox_1997}
\begin{equation} \label{comp}
\sum_{mn}\chi_{mn}B_n\dg B_m =\openone.
\end{equation}

 Although the process matrix $\chi$ can be obtained by simply inverting the $\mathcal{E}(\rho)$ determined from the measurements, there is no guarantee that errors in the measurement will lead to a process matrix which is completely positive and hence physical. A maximum likelihood estimation similar to that presented in Ref.\,\cite{obrien_QPT_2004} can be used. Here, we write the process matrix in a Cholesky decomposition of the form \begin{equation}
 \chi(\vec{t})= {T\dg T},
\end{equation} where $T$ is a lower triangular matrix parametrized by the vector $\vec{t}$. This ensures that $\chi$ be Hermitian. Next, the measured data is fit to a physical state by minimizing the function
\begin{equation}
 f(\vec{t})=\sum_{a,b=1}^{d^2} \left[m_{ab}-\sum_{m,n=0}^{d^2-1}\chi_{mn}\mathrm{Tr}[M_b B_m\ket{\phi_a}\bra{\phi_a}B_n\dg]\right]^2.
\end{equation}  Here, $m_{ab}$ is the measured data for the case where the state $\ket{\phi_a}$ was prepared and the observable $M_b$ was measured. A Lagrange multiplier is then used to impose the completeness condition [Eq. \eqref{comp}], such that we find the minimum of the function 
\begin{equation}
 \begin{split}
 f(\vec{t})&=\sum_{a,b=1}^{d^2} \left[m_{ab}-\sum_{m,n=0}^{d^2-1}\chi_{mn}\mathrm{Tr}[M_b B_m\ket{\phi_a}\bra{\phi_a}B_n\dg]\right]^2\\&+\lambda \sum_{k=0}^{d^2-1}\left[\sum_{m,n=0}^{d^2-1}\chi_{mn}\mathrm{Tr}[B_mB_kB_n\dg]-\mathrm{Tr}[B_k]\right]^2
 \end{split}
\end{equation}  
to obtain the most probable completely positive $\chi$ matrix corresponding to the measured values.